\documentclass[aps,prd,twocolumn,nofootinbib,superscriptaddress]{revtex4-2}

\pdfoutput=1

\usepackage[utf8]{inputenc} 
\usepackage{amsmath}
\usepackage{amssymb}
\usepackage{xcolor}
\usepackage{graphicx}
\usepackage{dcolumn}
\usepackage{bm}
\usepackage{microtype}
\usepackage{threeparttable}
\usepackage{mathtools}
\usepackage[colorlinks=true]{hyperref}
\usepackage{physics}
\usepackage{comment}
\usepackage{soul}

\newcommand{\rf}[1]{Eq.~(\ref{#1})}

\newcommand{\rff}[1]{Fig.~\ref{#1}}
\newcommand{\rfc}[1]{Ref.~\cite{#1}}

\renewcommand{\d}{{\mathrm d}}

\newcommand{\E}{{\mathcal E}}

\newcommand{\be}{\begin{eqnarray}}
\newcommand{\ee}{\end{eqnarray}}

\newcommand{\rev}[1]{\textcolor{black}{#1}}

\begin{document}

\title{Experimentally accessible drive-induced attractor in a Fermi gas near unitarity} 

\author{Michal P. Heller}
\email{michal.p.heller@ugent.be}
\affiliation{Department of Physics and Astronomy, Ghent University, 9000 Ghent, Belgium}
\affiliation{Institute of Theoretical Physics and Mark Kac Center for Complex
Systems Research, Jagiellonian University, 30-348 Cracow, Poland}

\author{Clemens Werthmann}
\email{clemens.werthmann@ugent.be}
\affiliation{Department of Physics and Astronomy, Ghent University, 9000 Ghent, Belgium}

\begin{abstract}
\noindent Hydrodynamic attractors describe loss of sensitivity to initial
conditions. Their earliest, expansion-driven stage is distinct from the later relaxation-driven mechanism and central to the theoretical paradigm but hard to access in heavy-ion collision experiments. We show within a coupled energy–bulk-pressure model that a
near-unitary Fermi gas driven by a rapid scattering-length sweep loses sensitivity to one state-space direction before bulk relaxation sets in -- visible only in the joint state
space, not as a universal single-variable curve. We outline an experimental $^{40}$K protocol based on
time-resolved contact measurements and a varied-ramp comparison.
\end{abstract}

\maketitle

\noindent \textbf{\emph{Introduction.--}}
The evident applicability of hydrodynamics to systems that are neither large nor close to equilibrium is a central question in non-equilibrium physics. It is especially pronounced in ultrarelativistic nuclear collisions, where matter created on spatio-temporal scales of a few $\mathrm{fm}$ is nevertheless described successfully by relativistic viscous hydrodynamics~\cite{Busza:2018rrf,Schenke:2021mxx}. Hydrodynamic attractors address this puzzle by identifying dynamical memory-loss mechanisms: initially distinct far-from-equilibrium states can approach a common evolution before local equilibrium is reached~\cite{Soloviev:2021lhs,Jankowski:2023fdz}. Establishing where this memory loss comes from, and whether it can be isolated experimentally, is the question we address here.

We will organize this approach to an attractor into three regimes and introduce a shorthand notation for them. The earliest regime, which we call \underline{stage~E(xpansion)/D(rive)}, is dominated by the external expansion or drive itself: in heavy-ion-motivated models, rapid longitudinal expansion separates excitations according to momentum and efficiently reduces sensitivity to the initial state~\cite{Blaizot:2017ucy}. The next regime, \underline{stage~T(ransients)}, is controlled by interaction-driven relaxation: non-hydrodynamic excitations decay exponentially, bringing the system toward near-equilibrium hydrodynamics~\cite{Heller:2021yjh,Chesler:2009cy,Heller:2014wfa,Du:2022bel}. The final regime, \underline{stage~H(ydrodynamics)}, is the ordinary hydrodynamic tail: gradients are weak, non-equilibrium pressure corrections are small, and Navier--Stokes constitutive relations governed by a few transport coefficients apply~\cite{Florkowski:2017olj}. This terminology is introduced here only as a diagnostic classification of mechanisms; in particular, stage~E/D is not guaranteed to occur in every microscopic theory and may well be absent in very strongly coupled systems~\cite{Kurkela:2019set}. Precisely for this reason, it is the most important regime to seek in a controlled experiment.

Most theoretical insight into hydrodynamic attractors has come from Bjorken flow~\cite{Bjorken:1982qr}, an effectively 0+1-dimensional model of nuclear-collision dynamics assuming transverse homogeneity as well as invariance under boosts in longitudinal direction. In this model, the dominant early-time effect is longitudinal expansion. Although many studies further assume conformal symmetry, the attractor picture has been extended to include transverse expansion~\cite{Romatschke:2017acs,Ambrus:2021sjg,Ambrus:2021fej,Ambrus:2022koq,An:2023yfq}, broken conformal symmetry~\cite{Romatschke:2017acs,Spalinski:2025ngd}, and other expanding geometries~\cite{Mitra:2024zfy}. Heavy-ion collisions themselves, however, provide only indirect access to the earliest part of the evolution, which is subsequently masked by hydrodynamic expansion, hadronization, and final-state interactions. A tabletop realization would therefore offer a qualitatively new way to test the expansion-driven part of the attractor paradigm.

Cold atomic gases provide such a route. They have already realized far-from-equilibrium phenomena motivated by high-energy physics, including non-thermal fixed points and collective flow~\cite{Berges:2020fwq,Ueda:2020ehs,Prufer:2018hto,Erne:2018gmz,Brandstetter:2023jsy,Li:2025}. A particularly close analogue is a homogeneous two-component Fermi gas close to unitarity, in which the scattering length $a$ can be tuned in time via an external magnetic field. Since $a(t)$ is the only quantity breaking scale invariance, varying it acts analogously to a homogeneous expansion drive~\cite{Son:2005rv,Fujii:2018aik}. In this setting the non-equilibrium contribution to the isotropic pressure, the bulk pressure $\Pi$, obeys~\cite{Fujii:2024yce}
\begin{equation}
\label{eq:pievo}
    \tau_\Pi\dot\Pi=-\Pi-\zeta\, V[a(t)] ,
\end{equation}
where $\tau_{\Pi}$ is the relaxation time, $\zeta$ the bulk viscosity, and
\begin{equation}
    V[a(t)]=3a(t)\partial_ta^{-1}(t)\,.
\end{equation}
Equation~\eqref{eq:pievo} already contains the two later regimes introduced above: exponential relaxation of transients when $\tau_\Pi\dot\Pi\simeq-\Pi$, and the Navier--Stokes tail when $\Pi\simeq-\zeta V[a(t)]$. The open question is whether the same platform also realizes a genuinely expansion-/drive-induced attractor regime.

We examine this possibility in a system where the bulk pressure relaxation is coupled to the evolution of the energy density $\E$. $\E$ follows an evolution equation according to Tan's contact relation, which quantifies the heating induced by the changing scattering length in terms of the contact density~\cite{TAN20082971,Fujii:2018aik} composed of equilibrium and non-equilibrium contributions. Since the bulk pressure is also determined from Tan's contact, an experimental verification of these dynamics requires only one time-resolved measurement channel. The early-time equations for $\mathcal{E}$ and ${\Pi}$ contain competing drive-induced terms. When the time dependence is re-expressed in terms of the inverse interaction parameter $u=(k_Fa)^{-1}$, it becomes apparent that the relaxation term is negligible roughly when $|\dot u|\gg \tau_\Pi^{-1}$. Then, the dynamics is controlled by the interval swept in $u$, rather than by the detailed functional form of $a(t)$. This identifies a drive-induced memory-loss mechanism in the near-unitary Fermi gas, in direct analogy with stage~E/D as defined above. \rev{This mechanism is physically distinct from the attractor realized in Refs.~\cite{Fujii:2024yce,Mazeliauskas:2025jyi}: there, the drive was tuned to both compete with relaxation and excite hydrodynamic modes, and the attractor that emerged was the late-time, equilibration-driven one. Here the drive alone erases memory at earlier times, providing a second, independent information-loss channel rather than an extension of the previous result.}

To assess whether this mechanism could be observed, we employ a phenomenological model with state dependent equilibrium transport coefficients. We use temperature- and scattering-length-dependent input for $\zeta$, $\tau_\Pi$, and the equilibrium contact density from Luttinger--Ward calculations, together with a normal-phase equation of state~\cite{Enss:2010qh,Enss:2019ydh,Ku_2012}. Because of the more nontrivial coupling of the non-equilibrium sector to the energy, the attractor is more subtle than the one-variable Bjorken-flow picture. In particular, trajectories need not collapse onto a single universal curve in the ratio $f_\Pi(t)=\Pi/\mathcal{E}$. Instead, universal behavior appears as the decreasing importance of one direction in the two-dimensional state space $(\E,\Pi)$, referred to as dimensional reduction~\cite{Heller:2020anv,Spalinski:2025ngd}.

We quantify this dimensional reduction by evolving localized ensembles of initial conditions and applying a principal-component analysis to snapshots of their trajectory sets in $(\E,\Pi)$ space. The resulting variance ratio separates the three regimes introduced above: an initial drive-induced fall, an intermediate exponential decay of transients, and a late Navier--Stokes tail. This provides a concrete diagnostic for future cold-atom searches for hydrodynamic attractors and shows that the experimentally relevant signature of the early-time attractor is apparent in the joint evolution of bulk pressure and energy density, not either observable alone.

For an experimental verification of stage E/D, $^{40}$K is a convenient platform. Its s-wave Feshbach resonance at $B_0=202\,$G is neither very broad nor very narrow~\cite{PhysRevA.72.013610}, so it allows for a realization of the necessary drive speeds of the scattering length, $\partial_t |k_Fa|^{-1}\gg\tau_\Pi^{-1}$, while still exhibiting universal scattering length physics near the resonance. With the broad resonance of $^6$Li, some combination of higher magnetic field ramp speeds and lower densities would be necessary to achieve the desired $\partial_t(k_Fa)^{-1}$. Homogeneity of the system can be achieved through a box potential. A feasibility estimate with details of a possible experimental setup is given in the End Matter.

Throughout this paper, we set $\hbar=k_B=1$.

\vspace{10 pt}

\noindent \textbf{\emph{Evolution equations for the near-unitary Fermi~gas.--}}
The bulk-pressure equation~\eqref{eq:pievo} is our first dynamical input. It has the same relaxation structure as the M\"uller--Israel--Stewart equations used to model viscous stresses in relativistic fluids~\cite{Muller:1967zza,Israel:1979wp,Muronga:2003ta}, and such equations provided the setting in which hydrodynamic attractors were first identified in Bjorken flow~\cite{Heller:2015dha}. In the cold-atom context, Eq.~\eqref{eq:pievo} was used in Refs.~\cite{Fujii:2024yce,Mazeliauskas:2025jyi} to study attractor dynamics generated by a time-dependent scattering length $a(t)$. However, there the transport coefficients were assumed to depend only on $a$. Describing the earlier regime of a stronger drive and larger change in energy density makes it necessary to consider the additional dependence on temperature. Including this effect is important for assessing how the attractor would appear in a realistic cold-atom setting.

The second essential dynamical input is the energy density~$\E$. Tan's contact relation gives~\cite{TAN20082971}
\begin{align}
\label{eq.dEdt}
    \dot \E=-\frac{C(t)}{4\pi m}\partial_t(a^{-1}),
\end{align}
where $C(t)$ is the full contact density~\cite{Fujii:2018aik}
\begin{align}
\label{eq.fullcontact}
    C(t)=C_{\rm eq}(t)+12\pi m\Pi(t) a(t)
\end{align}
containing both the equilibrium contact and the non-equilibrium contribution proportional to the bulk pressure. The latter couples the energy injected by the scattering-length drive back to $\Pi$, turning Eq.~\eqref{eq:pievo} into a genuinely coupled evolution problem in the $(\E,\Pi)$ state space.

We work close to unitarity, $|k_Fa|\gg1$ with $k_F$ being the Fermi momentum, which is reflected in the values of the transport coefficients we use. We express all quantities in their natural cold-atom units. Temperatures are written as $\theta=T/T_F$, with $T_F=E_F=k_F^2/(2m)$. In dynamical equations, the scattering length is expressed in terms of $u=(k_Fa)^{-1}$, while the energy density and related quantities are explicitly given in units of $E_F n$, where $n=k_F^3/(3\pi^2)$. We will use these conventions throughout the rest of this work.

To close the set of Eqs.~\eqref{eq:pievo} and~\eqref{eq.dEdt}, we must specify $\zeta$, $\tau_\Pi$, and $C_{\rm eq}$. Since the energy density is now dynamical, its influence on the transport coefficients also has to be taken into account. Although the strong drive may push the system far from equilibrium, we use equilibrium values of the bulk viscosity and relaxation time as input. This keeps the description directly tied to existing microscopic calculations while retaining the non-equilibrium dynamics generated by Eqs.~\eqref{eq:pievo} and~\eqref{eq.dEdt}. We describe the temperature dependence of $\zeta$, $\tau_\Pi$, and $C_{\rm eq}$ using interpolations and fits to Luttinger--Ward calculations~\cite{Enss:2010qh,Enss:2019ydh}.  In terms of the dependence on $a$, near unitarity, the bulk viscosity scales as $\zeta\sim a^{-2}$~\cite{Dusling:2013sea,Enss:2019ydh,Hofmann:2019jcj,NISHIDA2019167949}. For more details on the transport coefficients see the Supplemental Material.

Since we describe the energy density $\E$ as the dynamical variable while the transport input is naturally given as a function of temperature, we need an equation of state. As the scattering-length drive will heat up the gas during the evolution as described by Eq.~\eqref{eq.dEdt}, we focus on the normal phase, $\theta\gtrsim0.2$. This regime also simplifies both the theoretical treatment and possible experimental realization. In the normal phase, the chemical potential $\mu$ and energy density $\E$ are well approximated by those of a shifted free Fermi gas~\cite{Ku_2012}, which fixes the equation of state $\E(\theta)$, or equivalently $\theta(\E)$, as described in the Supplemental Material.

The relaxation time is fixed by the low-frequency bulk viscous response. Equation~\eqref{eq:pievo} corresponds to a Drude peak of the form~\cite{Fujii:2024yce}
\begin{align}
    \zeta(\omega)=\frac{i\chi}{\omega+i\tau_\Pi^{-1}}.\label{eq:drude}
\end{align}
The standard near-equilibrium bulk viscosity is then 
\begin{equation}
\label{eq.bulkviscosity}
\zeta \equiv \chi \, \tau_\Pi\;,
\end{equation}
which we use to determine $\tau_\Pi$ from the bulk viscosity $\zeta$ and the sum rule $\chi$ computed in Ref.~\cite{Enss:2019ydh}.

Strictly speaking, Eq.~\eqref{eq:drude} describes only the low frequency behavior of the bulk viscosity, which gives way to a $\omega^{-3/2}$-tail. We work here with the dynamically simplest model to qualitatively examine the possibility of drive-induced memory loss in cold atoms.

We measure time in units of $T_F^{-1}$. This choice is convenient because the microscopic relaxation time is dynamical, with $\tau_\Pi\sim T^{-1}$. Under the rescaling $t\to T_F^{-1}\,t$, all terms except the explicit $\Pi$ term acquire a factor of $T_F$, so Eq.~\eqref{eq:pievo} becomes
\begin{align}
    \dot\Pi&=- (T_F\tau_\Pi[\E(t)])^{-1} \Pi-\chi[\E(t),a(t)] V[a(t)] \;. \label{eq:pievo_rescaled}
\end{align}

Finally, following Ref.~\cite{Fujii:2024yce}, we choose a power-law drive for the inverse scattering length,
\begin{align}
\label{eq.formofexp}
    a^{-1}(t)=a_0^{-1}\,t^{-\alpha}\;,\quad \alpha>0,
\end{align}
which gives $V[a(t)]\propto t^{-1}$ in explicit analogy to Bjorken flow. The analysis below shows, however, that the drive-dominated attractor regime does not rely on this particular functional form. Its strength is controlled by the interval swept in $u=(k_Fa)^{-1}$, while the subsequent transient decay is largely insensitive to the drive except near the crossover to the Navier--Stokes tail. As the attractor behavior favors positive bulk pressure $\Pi$, we choose positive $a$ to avoid a negative and therefore unphysical contact (see Eq.~\eqref{eq.fullcontact}). In particular, unless stated otherwise, we choose $u(t_0)=0.2$.

\vspace{10 pt}

\begin{figure}[t]
    \centering
    \includegraphics[width=.86\columnwidth]{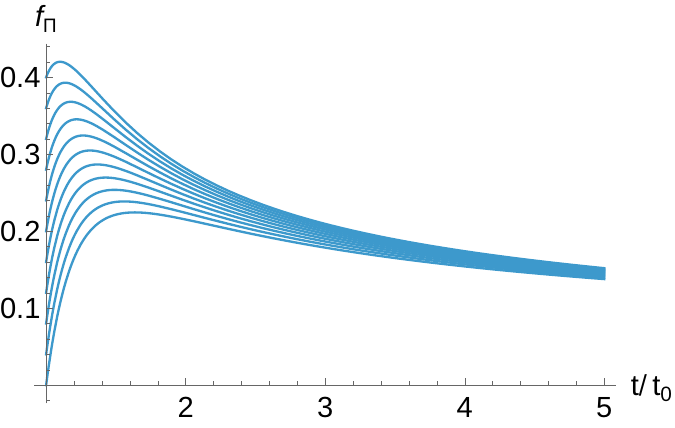}
    \includegraphics[width=.86\columnwidth]{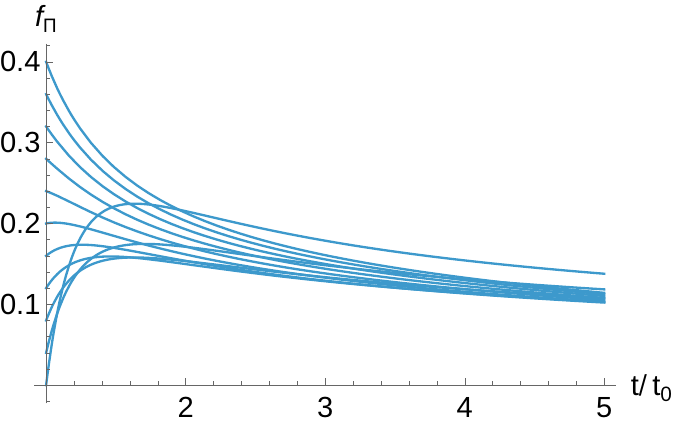}
    \caption{One-variable projection of the evolution in our model description. Curves show $f_\Pi(t)$ from the coupled evolution equations \eqref{eq:fevo} and~\eqref{eq.dEdt} for $\alpha=1$ and $f_\Pi(t_0)=0.04j$  with integer $0\le j\le10$. Top: fixed initial energy density $\E(t_0)=0.4E_Fn$. Bottom: varying initial energy density $\E(t_0)=E_Fn(0.4+0.2j)$. The drive starts at $t_0=10^{-2}$ with $k_Fa=1$ to magnify effects for illustrative purposes, so results are schematic. A common profile seems to appear only in the fixed-$\E(t_0)$ projection; varying $\E(t_0)$ shows that the attractor is not a single~curve~in~$f_\Pi(t)$. }
    \label{fig:earlyfatt}
\end{figure}

\noindent \textbf{\emph{Hydrodynamic attractors as fixed points for cold atoms.--}}
In analogy to Bjorken flow, the natural candidate for attractor behavior
in a single observable is the dimensionless bulk pressure
\begin{equation}
    f_\Pi \equiv \frac{\Pi}{\E}\,.
\end{equation}
The additional term
$-f_\Pi\E^{-1}\dot\E$ in its evolution equation introduces through the non-equilibrium part of the contact density in Eq.~\eqref{eq.fullcontact} a term that competes with the direct bulk-stress source at early times. This may lead to an early-time fixed point, a manifest property of the early-time Bjorken flow attractor, as explained in the Supplemental Material. The evolution equation for $f_\Pi$ becomes
\begin{align}
    \dot f_\Pi=3\left(f_\Pi^2
    -\chi/\E\right)u^{-1}\dot u -\tau_\Pi^{-1}f_\Pi\label{eq:fevo}\\
    +\frac{3\pi}{2}\frac{C_{\rm eq}/k_F^4}{\E/E_Fn}f_\Pi\dot u\;.\nonumber
\end{align}

Assuming no additional time dependencies, the $u^{-1}\dot{u}$-term would have two fixed points at $f_\Pi=\pm\sqrt{\chi/\E}$, of which one is attractive and the other repulsive. The attractive fixed point would constitute the early-time attractor. However, in the cold-atom case, the term $\chi/\E$ is strongly time dependent, which means that there is no early-time fixed point and beyond that no attractor behavior in a single observable. This motivates the diagnostics of dimensional reduction, that we will introduce in a moment.

Fig.~\ref{fig:earlyfatt} illustrates why tracking the observable $f_\Pi(t)$ alone is insufficient in the realistic case. For fixed initial energy density, the curves do appear to approach a common early-time profile. When the initial energy density is varied, however, no single universal curve remains. This is not a failure of attraction. Rather, the attractive value itself depends on $\E$. As $\E$ will also evolve differently for different values of $f_\Pi$, technically not even the curves for fixed initial $\E$ fully collapse. The realistic attractor is therefore not a curve in $f_\Pi(t)$, but a surface in the coupled $(\E,\Pi,t)$ state space. How much $\E$ changes throughout the evolution depends strongly on the initial size of $\Pi$ and the strength of the drive. Starting at $t_0=10^{-2}$, we saw an increase by $0.11\,E_Fn$ for $\alpha=1/8$ and $\Pi(t_0)=0.03\, E_Fn$, compared to $3.5\,E_Fn$ for $\alpha=1$ and $\Pi(t_0)=0.4\, E_Fn$.

\vspace{10 pt}

\noindent \textbf{\emph{Realistic cold-atom attractor.--}}
The ideal fixed-point structure described above is modified once the transport input appropriate to a near-unitary Fermi gas is used. The main difference is not the absence of an attractor, but a change in its form. With realistic transport coefficients, $\chi/\E$ is not constant: close to unitarity $\chi\propto a^{-2}$, and $\chi$ also depends nonlinearly on the evolving energy density. In addition, the equilibrium contribution to the contact density introduces a further drive-dependent term. Consequently, the values $f_\Pi=\pm\sqrt{\chi/\E}$ are no longer exact fixed points of Eq.~\eqref{eq:fevo}. They should instead be viewed as instantaneous reference values toward which the drive-dominated dynamics tend to move the system. Before introducing the diagnostics of memory loss, we first discuss what behavior can be expected. 

Let us examine the drive-dominated regime more closely by rewriting the evolution equations. For a monotonic drive of $a(t)$, the inverse interaction parameter $u=(k_Fa)^{-1}$ is in one-to-one correspondence with time, so we perform a substitution $f_\Pi(t)=\tilde f_\Pi(u(t))$ and $\E(t)=\tilde \E(u(t))$. The coupled equations become
\begin{align}
    \partial_u\tilde{f}_\Pi&=3\frac{\tilde{f}_\Pi^2}{u}-3\frac{\chi[\tilde{\E},u]}{\tilde{\E}u}+\frac{3\pi}{2}\frac{C_{\rm eq}[\tilde{\E}]/k_F^4}{\tilde{\E}/E_Fn}\tilde{f}_\Pi\nonumber\\
    &\quad-(\tau_\Pi\dot u)^{-1}\tilde{f}_\Pi\;,\label{eq:ufevo}\\
    \frac{\partial_u\tilde{\E}}{E_Fn}&=-3\frac{\tilde{f}_\Pi \tilde{\E}}{u~E_Fn}-\frac{3\pi}{2}{C_{\rm eq}[\tilde{\E}]/k_F^4}\;.\label{eq:uEevo}
\end{align}

We define drive-dominated memory loss as the early part of memory loss that is identical to the case where the relaxation term $-(\tau_\Pi\dot{u})^{-1}f_\Pi$ is dropped from the dynamical equation. In this case the equations are independent of the detailed functional form of $a^{-1}(t)$; the amount of early-time memory loss is controlled instead by the interval swept in $u$. This makes the drive-dominated regime robust against changes in the drive protocol. The relaxation term need not be small compared to the others, since memory loss as a property of neighboring trajectories can still be similar even if individual trajectories deviate significantly, which we do observe. Still, $|\dot u|\gg\tau_\Pi^{-1}$ at $t_0$ provides a rough criterion for the necessary drive speed to achieve drive dominated memory loss. 

The relevant hierarchy appears experimentally plausible. Since $\tau_\Pi |\dot u|\sim k_F^{-3}\sim E_F^{-3/2}$, lower densities are preferable. At $E_F=h\times1\,$kHz, we find a relaxation time of $\tau_\Pi\sim0.033\,{\rm ms}$, while driving rates close to a Feshbach resonance can reasonably reach $\partial_t|k_Fa|^{-1}\sim550\,{\rm kHz}$ (cf. End Matter). Thus the drive-dominated regime in these setups can extend up to $|\dot u|\sim 18\,\tau_\Pi^{-1}$, leaving room for an experimentally resolvable separation between drive-dominated and relaxation-dominated dynamics. \rev{A discussion of feasible realization in a uniform $^{40}$K gas---including the magnetic-field drive, the protocol for reconstructing $\E$ and $\Pi$, and the relevant experimental limitations---is given in the End Matter.}

At later times, the relaxation term in Eq.~\eqref{eq:ufevo} takes over. The evolution then crosses into the transient regime, $\dot f_\Pi\simeq-\tau_\Pi^{-1}f_\Pi$, and eventually reaches the Navier--Stokes tail, $f_\Pi\simeq -3\zeta \E^{-1}u^{-1}\dot u$. Since $\dot u$ is small in this late regime, the energy density changes slowly and the tail scales with $u\dot u$. While this discussion neglects the equilibrium contact term, in practice we find that it has little effect on the qualitative behavior.

\begin{figure}[t]
    \centering
    \includegraphics[width=\columnwidth]{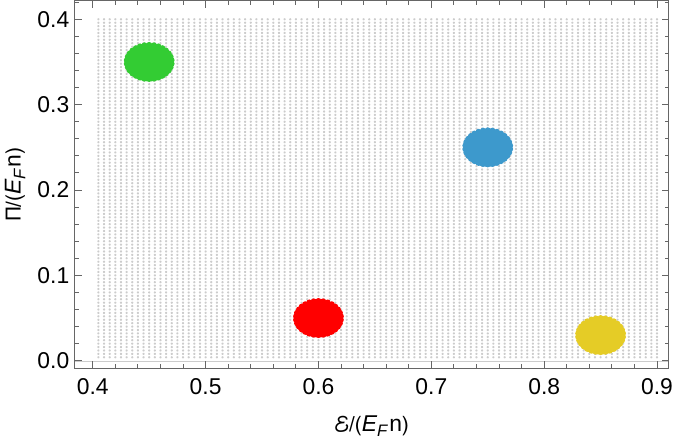}
    \includegraphics[width=\columnwidth]{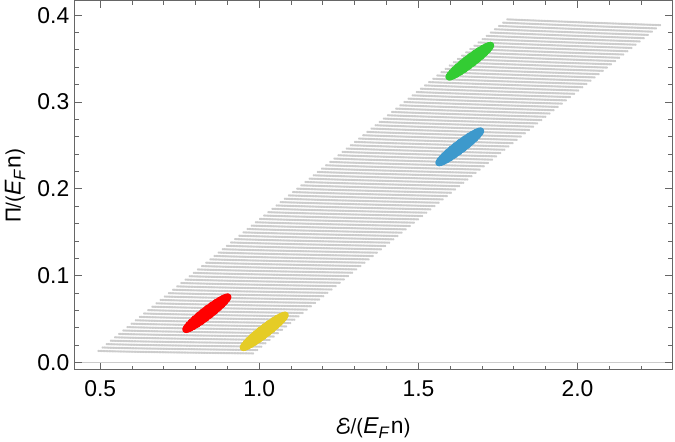}
    \includegraphics[width=\columnwidth]{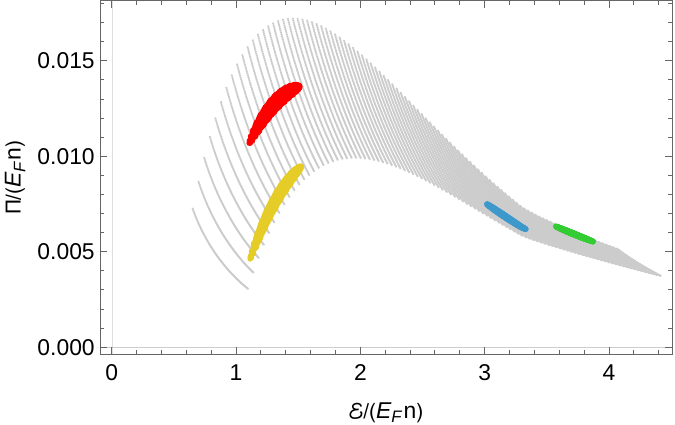}
    \caption{Dimensional reduction in the model evolution. The gray points show an initially rectangular set of $10\,000$ solutions, and the colored points show four initially spherical ensembles of 465 solutions used in the principal-component analysis. All solutions are evolved with $\alpha=1$ in the $(\E,\Pi)$ plane and shown at $t=t_0=10^{-2}$ (top), $t=3t_0$ (middle), and $t=1$ (bottom). Note the scales of the axes. The flattening of the ensemble is the state-space signature of the attractor.}
    \label{fig:setatt}
\end{figure}

\noindent \textbf{\emph{Dimensional reduction.--}}
The coupled nature of the realistic attractor suggests an adapted diagnostic: instead of asking whether many solutions collapse onto a single curve in $f_\Pi(t)$, we ask whether sets of initial conditions lose sensitivity to a direction in the $(\E,\Pi)$ plane. Figure~\ref{fig:setatt} shows this dimensional reduction for solutions of Eqs.~\eqref{eq.dEdt} and~\eqref{eq:fevo}. An initially rectangular set of $10\,000$ points is rapidly deformed into a thin band. Thus the realistic cold-atom attractor appears as directional loss of memory in state space.

This behavior is closely related to the geometric picture of hydrodynamic attractors developed for Bjorken flow~\cite{Heller:2020anv}, but with an important cold-atom difference. Here, the direction of lesser importance changes with time and depends on the location in the $(\E,\Pi)$ plane, because the transport input and the equation of state evolve along each trajectory. The state-space diagnostic is therefore not only a visualization tool; it is the natural way to define the attractor in the general case.

To quantify the dimensional reduction, we follow Ref.~\cite{Heller:2020anv} and perform a principal-component analysis on localized ensembles of solutions. We initialize spherical sets of 465 configurations in the $(\E,\Pi)$ plane with a radius of $0.02\,E_Fn$ -- shown in Fig.~\ref{fig:setatt} in red, blue, green and gold -- so that both directions are measured in the same units and have equal initial variance. Let
\begin{align}
R_{\rm PCA}(t)=\frac{\lambda_2(t)}{\lambda_1(t)}
\end{align}
denote the ratio of the smaller principal variance to the larger one. A decrease of $R_{\rm PCA}$ measures the decreasing importance and eventually effective loss of one state-space direction and therefore gives a local measure of attractor formation. Note that $R_{\rm PCA}$ may decrease also through an increase in $\lambda_1$, which nevertheless also means that the direction of $\lambda_2$ becomes less important. Indeed, if the orientation of the associated state space vectors remains constant, the ratio of state projections onto these directions would show a collapse to a universal curve, which would unambiguously be identified as attractor behavior.

\begin{figure}[t]
    \centering
    \includegraphics[width=\columnwidth]{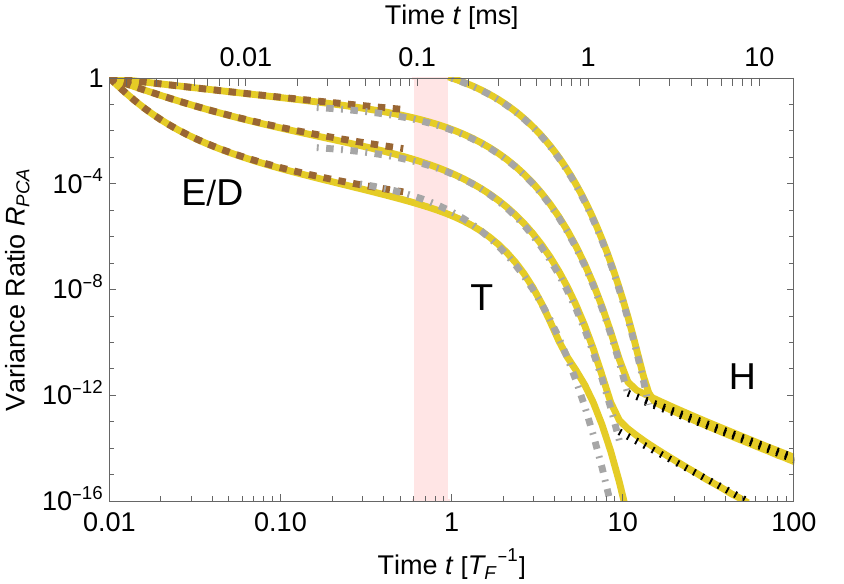}
    \caption{Quantifying dimensional reduction. Shown is the ratio $R_{\rm PCA}=\lambda_2(t)/\lambda_1(t)$ of the subleading to the leading principal variance for localized ensembles of solutions. As indicated by the color, results are for the gold set of initial states in Fig.~\ref{fig:setatt}, though other sets give similar results. The curves correspond from top to bottom to $\alpha=1/8$ and $t_0=1$, $\alpha=1/8$ and $t_0=10^{-2}$, $\alpha=3/8$ and $t_0=10^{-2}$, and $\alpha=1$ and $t_0=10^{-2}$. They are compared with the same drive-only evolution from Eqs.~\eqref{eq:ufevo} and~\eqref{eq:uEevo} without relaxation, re-expressed in $t$-evolution according to three different drives (``E/D", brown, dashed, absent for $t_0=1$ relaxation-dominated reference curve), transient exponential relaxation (``T", gray, dot-dashed), and the Navier--Stokes tail (``H", black, dotted). The same diagnostic separates drive-induced memory loss, transient decay, and hydrodynamic response. The decay of $R_{\rm PCA}$ before the exponential trend, which depends only on the interval swept in $u=(k_Fa)^{-1}$ as indicated by agreement with the brown dashed line, is the falsifiable prediction of drive-induced memory loss. Time is given in units of $T_F^{-1}$ on the bottom axis and in units of ms for a gas at $E_F=h\times1\,$kHz on the top axis. The light red band shows the range of values of $\tau_\Pi$ for the four heating histories. Earlier start or stronger drive means more heating and shorter $\tau_\Pi$. The $\alpha=1$ curve deviates from the $\propto [\exp(-t/\tau_\Pi)]^2$-trend at $tT_F\sim4$, only to return to it again shortly after. This coincides with a very fast transition of the dominant term in the contact from non-equilibrium to equilibrium, which is unique to this case. }
    \label{fig:pcr}
\end{figure}

The evolution of $R_{\rm PCA}$, shown in Fig.~\ref{fig:pcr}, resolves the three regimes introduced in the Introduction. At early times, $R_{\rm PCA}$ follows the result obtained from Eqs.~\eqref{eq:ufevo} and~\eqref{eq:uEevo} after dropping the relaxation term. The same reference evolution was used in all three cases, demonstrating independence of the form of the drive in stage E/D, as can also be seen more directly in Fig.~\ref{fig:u-independence} in the Supplemental Material. This is the drive-dominated regime and is the central result of this work. At intermediate times, the ratio decays as the square of the transient relaxation, $R_{\rm PCA}\sim [\exp(-t/\tau_\Pi)]^2$. At late times, the solutions reach the Navier--Stokes tail and the variance ratio follows the corresponding scaling, $R_{\rm PCA}\sim [u(t)\dot u(t)]^2$.

Several features make this diagnostic experimentally useful. First, the early-time fall agrees with the reduced drive-only equations, showing that the dimensional reduction is not caused by equilibration and is sensitive only to the swept interval in the scattering-length drive. Second, the intermediate relaxation regime is reliably identified by exponential decay and is largely independent of the drive. Third, the late-time tail has the expected Navier--Stokes scaling. The same observable therefore distinguishes drive-induced memory loss, transient relaxation, and hydrodynamic response within a single cold-atom protocol.

\vspace{10 pt}

\noindent \textbf{\emph{Conclusion and outlook.--}}
We have identified a cold-atom realization of the expansion-driven part of hydrodynamic-attractor dynamics. In heavy-ion collisions, this early memory-loss mechanism is difficult to isolate because it is followed by hydrodynamic expansion, hadronization, and final-state interactions; final-state observables therefore provide only indirect access to the earliest times~\cite{Giacalone:2019ldn}. A near-unitary Fermi gas with a time-dependent scattering length avoids this limitation: the drive $a(t)$ can be controlled directly, and the ensuing evolution can be probed before microscopic relaxation fully dominates.

The central result is that the expansion-/drive-induced regime emerges once the energy density $\E$ is tracked together with the bulk pressure $\Pi$. Through Tan's contact relation, the changing scattering length couples the drive-induced heating to the non-equilibrium pressure, whose evolution is coupled back to the energy density via the temperature dependence of transport coefficients, producing state-dependent attractive behavior. The principal-component diagnostic in Fig.~\ref{fig:pcr} resolves the sequence introduced above: an initial drive-dominated stage, an intermediate transient exponential decay, and a late Navier--Stokes tail. This shows that cold atoms can realize not only the relaxation-driven attractor discussed in Ref.~\cite{Fujii:2024yce}, but also the early-time attractor mechanism generated by the drive itself. \rev{The two are physically distinct: the early-time channel removes memory through the drive, before equilibration, and is therefore not a refinement of the relaxation-driven case but a separate mechanism.}

A key practical consequence is that the realistic attractor is not a universal curve in a single observable such as $f_\Pi(t)=\Pi/\E$. \rev{In particular, the single-curve diagnostic that identifies the equilibration-driven attractor of Ref.~\cite{Fujii:2024yce} does not apply here.} With temperature- and scattering-length-dependent transport input, the attractive structure appears instead as loss of sensitivity to one direction in the coupled $(\E,\Pi)$ state space, as shown in Fig.~\ref{fig:setatt}. Future searches should therefore target the joint evolution of bulk pressure and energy density. This provides a concrete criterion for distinguishing drive-induced memory loss from equilibration-driven relaxation in cold-atom experiments.

The next step is to adapt this diagnostic to periodic modulations of the scattering length, which are natural in experiments and have already been explored in related attractor proposals~\cite{Mazeliauskas:2025jyi}. More broadly, the framework developed here turns hydrodynamic attractors from an indirect feature of nuclear-collision phenomenology into a controllable non-equilibrium phenomenon in a tabletop quantum gas. \rev{It thereby links the study of far-from-equilibrium dynamics in heavy-ion collisions and quantum field theory to controllable experiments with ultracold atoms, a connection recently exploited for collective-flow phenomena~\cite{Brandstetter:2023jsy,Li:2025}.}

In the present setup, we have made several choices in order to arrive at a set of simple equations for all considered quantities, such as dropping subleading terms in the expansion around the unitary limit in large scattering length $k_{F} a$ and considering the normal phase as opposed to a quantum condensate. Naturally, it will be interesting to study how the attractor behavior changes when going beyond these assumptions. The approximation most relevant to the off-equilibrium stage~E/D is that of considering static equilibrium transport coefficients. Non-equilibrium effects may introduce qualitatively new behavior to the early-time stage. For example, it is well known that the frequency dependent bulk viscosity exhibits beyond the Drude form at small frequencies also a large frequency tail $\sim \omega^{-3/2}$. This large frequency behavior dictates the dynamics on short timescales. Going even beyond that, alternative dynamical descriptions via kinetic theory or holography capture more microscopic aspects of the dynamics and are therefore intrinsically capable of describing far-from-equilibrium dynamics much more accurately. In that vein, universal dynamics in a cold atom system has recently been examined in a fully quantum description~\cite{Sharell:2026mrv}.

\vspace{10 pt}
\begin{acknowledgments}
It is a pleasure to thank Alex Buchel, Matisse De Lescluze, Gabriel Denicol, Yusuke Nishida and Alex Serantes for discussions and correspondence and Robbe Brants, Matisse De Lescluze and Micha{\l} Spali{\'n}ski for comments on the draft. We are grateful to Tilman Enss, Thomas Schäfer and Keisuke Fujii for helpful suggestions on how to improve the first manuscript and to Tilman Enss for providing data from the Luttinger-Ward model. We thank Martin Gažo for helpful advice with the proposed experimental setup. We also benefited from ChatGPT 5.5 and 5.6 and Claude 4.8 and 5.0 in preparing the latest version of the manuscript, in particular in detailing the experimental setup. This project has received funding from the European Research Council (ERC) under the European Union’s Horizon 2020 research and innovation programme (grant number: 101089093 / project acronym: High-TheQ). Views and opinions expressed are however those of the authors only and do not necessarily reflect those of the European Union or the European Research Council. Neither the European Union nor the granting authority can be held responsible for them. This work was also partially supported  by the Priority Research Area Digiworld under the program Excellence Initiative  - Research University at the Jagiellonian University in Krakow.

\end{acknowledgments}

\bibliography{literature_cold_attractors}

\section*{END MATTER: Experimental realization and measurement protocol}\label{app:exp}

The analysis in the main text shows that the drive-dominated regime (stage~E/D) is encoded in the \emph{joint} evolution of the energy density~$\E$ and the bulk pressure~$\Pi$ [Eqs. \eqref{eq.dEdt} and \eqref{eq:pievo_rescaled}], and that it becomes visible as the dimensional reduction of an ensemble of initial conditions in the $(\E,\Pi)$ plane [\rff{fig:setatt} and \rff{fig:pcr}]. Here we describe a concrete cold-atom setting in which this evolution can be prepared and read out.

\subsection*{Platform and drive}
A natural platform is a two-component gas of $^{40}$K prepared near the s-wave Feshbach resonance at $B_0\approx 202\,$G. Loading the gas into a uniform optical box potential~\cite{Mukherjee:2017} realizes the spatially homogeneous conditions assumed throughout the main text, eliminating the trap averaging and density gradients that would otherwise obscure a homogeneous system.

Near a Feshbach resonance the scattering length as a function of an external magnetic field follows the standard parametrization $a(B)=a_{\rm bg}\,[1-\Delta/(B-B_0)]$~\cite{Chin:2010}, so that any target protocol $a^{-1}(t)$ is realized by the magnetic-field ramp obtained by inverting this relation. The criterion for a universal scattering length in the system is $E_F\ll E_0$, where $E_0$ is the coupling energy scale, which depends on the properties of the resonance~\cite{Ketterle_2008}. In particular, $E_0$ would likely be too small for very narrow resonances. For the $202\,$G resonance of $^{40}$K, $E_0=h\times21.8\,$MHz~\cite{PhysRevA.72.013610}, which is much larger than $E_F$ at the small densities needed for drive domination. The power-law drive of \rf{eq.formofexp}, $a^{-1}(t)=a_0^{-1}t^{-\alpha}$, is one convenient choice; as established in the main text, however, the early-time memory loss depends only on the interval swept in $u=(k_Fa)^{-1}$ and not on the detailed functional form of $a(t)$ [cf. Eqs. \eqref{eq:ufevo} and \eqref{eq:uEevo}]. The protocol can therefore be adapted to whichever field ramp is most cleanly executed. 

A rough drive speed target is the timescale hierarchy $|\dot u|\gg\tau_\Pi^{-1}$ identified in the main text. The $202\,$G resonance in $^{40}$K gases features $a_{\rm bg}=174a_0$ and $\Delta=7.1\,$G~\cite{2010NatPh...6..569G,JPGaebler_PhD}. At $E_F=h\times1\,$kHz, or equivalently $k_F^{-1}=6800a_0$, the scattering length will thus be modulated at a rate of $\d|k_Fa|^{-1}/\d B=5.5\, \mathrm{G}^{-1}$. Field slews of order $100\,\frac{\mathrm{G}}{\mathrm{ms}}$ have been demonstrated in experiments with $^6$Li~\cite{PhysRevLett.92.120403,PhysRevA.71.045601}, so a similar speed may also be reached in a $^{40}$K experiment, though this would have to be verified. With that, the modulation rate in time is then $\partial_t|k_Fa|^{-1}\sim550\,$kHz. Since most of the heating of the system will occur at the early stage, leading to shorter relaxation times, for a conservative estimate we compute the relaxation time at $\theta=3$ according to Eq.~\eqref{eq:taupi} in the Supplemental Material and find $\tau_\Pi=0.2T_F^{-1}=0.033\,$ms. Note that this is among the highest temperatures we found, while typically they stay below $\theta=1$. With this, $|\dot u|\sim18\,\tau_\Pi^{-1}$, so the rough criterion for experimental accessibility of stage~E/D can be fulfilled.

\subsection*{Measurement protocol}
The diagnostic requires tracking the trajectory of the pair $(\E,\Pi)$, both components of which can be reconstructed experimentally.

The bulk pressure is obtained from the contact. Tan's contact density $C(t)$ has been measured in strongly interacting Fermi gases via radio-frequency- (rf-)induced dimer projection~\cite{44rc-34n1} or from the large-momentum $k^{-4}$ tail measured via momentum-resolved rf spectroscopy~\cite{Stewart:2010,Sagi:2012,Mukherjee:2019}. In the present protocol the challenging step is the time-resolved, non-equilibrium version of this measurement. It can be implemented as a destructive snapshot sequence: the same preparation and drive are repeated, the evolution is interrupted at different times and the contact is read out after a rapid projection. Given the measured $C$ and the equilibrium contact $C_{\rm eq}(\theta)$ fixed by the equation of state (see Supplemental Material), the bulk pressure is extracted from the contact relation~\rf{eq.fullcontact},
\begin{align}
    \Pi(t)=\frac{C(t)-C_{\rm eq}\!\left[\theta(t)\right]}{12\pi m\,a(t)}\,.
    \label{eq:Piextract}
\end{align}

The short timescales required for this experiment call for measurements with high time resolution. %, $t\lesssim \hbar / E_F$, direct measurements cannot resolve the energy because of quantum energy-time-uncertainty. Instead, 
The energy density $\E$ can be determined from the history of contact measurements by time integration of Tan's contact relation, Eq.~\eqref{eq.dEdt}. The contact %is measured via a destructive snapshot, and it is a high momentum observable, so it
can be read out from a destructive short pulse. Ref.~\cite{44rc-34n1} introduces a measurement of the contact via dimer projection and tests it for the $202\,$G resonance of $^{40}$K, explicitly mentioning hydrodynamic attractors as a possible application. In the End Matter of~\cite{44rc-34n1}, a lower bound on the pulse duration of the technique is stated as $t_{\rm rf}\ge1\,\mathrm{\mu s}$. This would make it possible to resolve $\mathcal{O}(10)$ contact measurements across the first $30\,\mathrm{\mu s}$, which is the scale on which stage E/D manifests itself in our theoretical description. A more fundamental question is the timescale on which the large-momentum tail adjusts to a change in $a$. In equilibrium it is set by short-range physics ($\sim \hbar/E_0$), but its behavior under fast nonequilibrium drives is unclear -- another reason to investigate this regime. What remains is to determine the initial energy $\E(t_0)$, i.e. the constant in the contact integration, which can be calibrated in equilibrium or for very slow drives. The equation of state is not measured during the fast non-equilibrium ramp, it enters only as an equilibrium calibration used to convert $\mathcal{E}$ to the reduced temperature $\theta$ and the equilibrium contact $C_{\rm eq}$, see Supplemental Material for the theory relations.

In order to reveal a dimensional reduction in $(\E,\Pi)$-space, the protocol has to be repeated several times for an ensemble of initial states. In order to prepare initial states with different $\E$ and $\Pi$, one may vary initial temperatures and apply a prior slow drive in the Navier--Stokes regime, where $\E$ and $\Pi$ behave in easily predictable ways. Fig.~\ref{fig:pcr} shows the behavior of $R_{\rm PCA}$ for the gold ensemble of initial states in Fig.~\ref{fig:setatt}, which features comparatively small bulk pressures $\Pi/\mathcal{E}\sim0.03$. The behavior is similar for the other sets, so it can be expected to remain unchanged for even smaller $\Pi$. After switching to the chosen fast driving protocol, the evolution of this ensemble is tracked in the $(\E,\Pi)$ plane and a principal component analysis is performed in order to measure the principal-component variance ratio $R_{\rm PCA}(t)$ of \rff{fig:pcr}. We stress that, in contrast to the relaxation-driven attractor of \rfc{Fujii:2024yce}, neither $\E$ nor $\Pi$ alone exhibits stage~E/D.

\subsection*{Predicted signal}

\rev{Fig.~\ref{fig:pcr} shows what to expect for the evolution of the dimensional reduction observable $R_{\rm PCA}$. The top axis shows time in laboratory units for a gas at $E_F=h\times1\,$kHz. Starting from a localized ensemble at reduced temperature $\theta\approx0.6$, the variance ratio collapses by several orders of magnitude within the first few tens of microseconds, i.e. on a timescale smaller than $\tau_\Pi$.}

The early stage~(E/D) coincides with the drive-only reference and is therefore insensitive to the relaxation time; the transition to the transient~(T) stage starts around $t\sim0.03\,$ms and finally the hydrodynamic tail~(H) appears, whose drive dependence is manifest in the markedly different late-time behavior of the two ramps. The exponential behavior of stage T and power-law behavior of stage H may be fitted appropriately, making any prior decrease that is unaccounted for a clear indication of drive-induced memory loss. Alternatively, one may repeat the entire measurement with a different driving protocol and compare the evolution of $R_{\rm PCA}$ as a function of $u=[k_Fa(t)]^{-1}$, in anticipation of independence of the precise form of $a(t)$ only in stage E/D as in our model description. Different drives will have an influence on the value of $\tau_\Pi$ through different amounts of heating. This is directly demonstrated in Fig.~\ref{fig:u-independence} in the Supplemental Material. Thus, the drive-induced memory loss would be revealed as the part of the evolution of $R_{\rm PCA}$ that is identical in the two cases. 

The speed of the drive may be adjusted to compromise between how pronounced the drive-dominated part is and experimental resolvability of $R_{\rm PCA}$. While precise predictions are model dependent, our hydrodynamic results give an idea of the order of magnitude. For $\alpha=1/8$, we see a drop by one decade during stage E/D, while for $\alpha=1$ it drops by four decades. For more accurate values, see Table~\ref{tab:alphacases}. Of course, the experimental verification of drive-induced memory loss does not need to resolve the entire progression through stage T and stage H, but only the distinction to transient decay driven memory loss that will set in in the transition from stage E/D to stage T.

\subsection*{Limitations}
Several effects constrain the protocol without precluding it. The system should be kept in the normal phase, $\theta\gtrsim0.2$, both to remain within the equation of state given in the Supplemental Material and to stay above the superfluid transition; because the drive heats the system, together with a potential upper limit on the temperature that the setup may require, this bounds the interval that can be swept in~$u$.

The contact readout is destructive and has to be repeated for several points in time -- to allow for the time integration -- and across the ensemble of initial states. Planning for $\mathcal{O}(10)$ initial conditions, measurement time points and repetitions for statistics leads to $\mathcal{O}(1000)$ total shots for the experiment. Integration helps with statistical errors by averaging, and common-mode systematic errors partially cancel in computing $R_{\rm PCA}$, since only distances in phase space matter. Thus, the resolution will likely be limited by the statistical noise (variance $\lambda_{\rm noise}$) in the pressure measurement via the contact. Resolving the narrowing of the ensemble in phase space will require an accuracy of $\sqrt{\lambda_{\rm noise}/\lambda_1}\lesssim \sqrt{\lambda_2/\lambda_1}$. The resulting values for our three example cases are compiled in Table~\ref{tab:alphacases} together with the swept range in $u$ and $B$. Larger $\lambda_2$ at the time of transition to transient decay is favorable in terms of the signal to noise ratio.

\begin{table}[]
    \centering
    \begin{tabular}{c||c|c|c|c}
       $\alpha$  & $t_{\rm rel}\,[\mathrm{\mu s}]$ &$\Delta u$ & $\Delta B\,[\mathrm{mG}]$ & $\sqrt{\lambda_2/\lambda_1}$ \\
       \hline
        1/8 & 29 & 0.061 & 11 & 0.34\\
        3/8 & 22 & 0.12 & 23 & 0.091\\
        1 & 19 & 0.18 & 33 & 0.017
    \end{tabular}
    \caption{Properties of stage E/D for the three considered drives: relaxation memory loss onset times $t_{\rm rel}$ defined as the first times that the full and drive-only results differ by more than 10\% and values of the total sweep in $u$ and $B$ at $k_F^{-1}=6800a_0$ and the necessary accuracy $\sqrt{\lambda_2/\lambda_1}$ at those times.}
    \label{tab:alphacases}
\end{table}

In terms of sufficient lifetime of the system, in a two-component Fermi gas atom loss near the resonance via three-body recombination is strongly suppressed by Pauli blocking~\cite{Petrov:2004,GiorginiPitaevskiiStringari:2008}. In practice, the lifetime of the system constituents near the resonance exceeds $100\,$ms~\cite{PhysRevLett.92.083201}. As Fig.~\ref{fig:pcr} shows, the entire evolution history proceeds within the system's lifetime.

Finally, the use of static, equilibrium transport coefficients is least controlled at the earliest times, where the large-frequency $\omega^{-3/2}$ tail of the bulk-viscosity spectral function---\rev{beyond the low-frequency Drude response underlying the relaxation equation~\rf{eq:pievo}}---can modify the dynamics; a fully quantitative early-time prediction would incorporate this frequency dependence. These caveats delimit the regime of validity rather than constitute an obstruction, placing the drive-dominated attractor within reach of present cold-atom capabilities.

\newpage

\clearpage

\newpage

\appendix

\setcounter{page}{1}

\begin{center}
 {\Large SUPPLEMENTAL MATERIAL}
\end{center}

\section{Hydrodynamic attractors as fixed points for cold atoms}
Hydrodynamic attractors in Bjorken flow can be understood as trajectories connecting fixed points that control different dynamical regimes~\cite{Blaizot:2017ucy,Blaizot:2019scw}. We now attempt to adapt this viewpoint to the near-unitary Fermi gas. The goal is to identify which variable, and which part of the evolution, can display memory loss driven by the scattering-length drive rather than by microscopic relaxation.

Consider first a single quantity $f(t)$ obeying a first-order equation
\begin{align}
    \partial_tf=G(f,t)\;.
\end{align}
If $G(f,t)$ can be organized into terms with different time dependences, then in a regime where one of these terms dominates the evolution equation reduces to
\begin{align}
    \partial_t f =T(t)F(f)\;,
\end{align}
where $T(t)$ describes this dominant time dependence and $F(f)$ contains the dependence on the dynamical variable. For positive $T(t)$, zeros of $F$ are fixed points of the reduced dynamics: they are attractive or repulsive depending on whether $F$ decreases or increases through the zero. In contrast, in the case of a constant $F$, all solutions change by the same amount over the course of the evolution and therefore their separation stays constant. This distinction will be important below. The same reasoning applies to coupled equations whenever the additional variables have a fixed time dependence in the regime of interest and can be absorbed into $T(t)$.

In Bjorken flow, expansion terms scale as $1/\tau$ and dominate at early proper time, while relaxation terms dominate later. This produces an early expansion-controlled repulsor--attractor pair and a late attractive fixed point associated with the decay of non-hydrodynamic stress~\cite{Heller:2015dha}. We now ask whether the cold-atom evolution contains the same structure.

Applying the argument to the bulk pressure $\Pi$ in Eq.~\eqref{eq:pievo}, we can conclude that considering only the evolution of $\Pi$ with a temperature-independent $\zeta$ is not sufficient for early-time attractor behavior. For the drive
$a^{-1}(t)=a_0^{-1}t^{-\alpha}$, the source term generated by the changing scattering length scales as $t^{-2\alpha-1}$ and is independent of $\Pi$, whereas the relaxation term is proportional to $\Pi$ and has no explicit drive enhancement. Thus, for sufficiently large $a_0^{-1}$, the early-time drive builds up bulk pressure almost independently of the initial value of $\Pi$, meaning different trajectories do not dynamically approach each other. The genuine attractor in $\Pi$ alone therefore emerges only in the later relaxation toward the Navier--Stokes value, $\Pi\simeq-\zeta V[a(t)]$, as emphasized in Ref.~\cite{Fujii:2024yce}. 

To expose attractor behavior facilitated by the drive in a single observable, one must consider a dynamical variable whose early-time evolution is itself state-dependent. The natural choice is the dimensionless bulk pressure
\begin{equation}
    f_\Pi \equiv \frac{\Pi}{\E}\,.
\end{equation}
Taking the time derivative of this ratio introduces the term
$-f_\Pi\E^{-1}\dot\E$, which couples the bulk-pressure dynamics to the heating induced by the time-dependent scattering length. Through the non-equilibrium part of the contact density in Eq.~\eqref{eq.fullcontact}, this term is nonlinear in $f_\Pi$ and can compete with the direct bulk-stress source as is evident from Eq.~\eqref{eq:fevo} in the main text, which we state again below. This is the cold-atom counterpart of the normalized stress variables used in analyses of the Bjorken-flow attractor~\cite{Heller:2011ju,Heller:2015dha}.

\begin{align}
    \dot f_\Pi=3\left(f_\Pi^2
    -\chi/\E\right)u^{-1}\dot u -\tau_\Pi^{-1}f_\Pi\label{eq:fevo_app}\\
    +\frac{3\pi}{2}\frac{C_{\rm eq}/k_F^4}{\E/E_Fn}f_\Pi\dot u\;.\nonumber
\end{align}

The mechanism is clearest in a limiting case that removes inessential time dependencies. Suppose that $\tau_\Pi$ has the same scattering-length dependence as $\zeta$, so that $\chi/\E$ is independent of $a$ and only weakly dependent on $\E$, as is the case for example in kinetic theory of massive particles~\cite{DeGroot:1980dk} and in non-conformal holographic models~\cite{Buchel:2015ofa}. Then the bulk-stress source and the non-equilibrium contact contribution to $\dot\E$ carry the same explicit drive factor, $u^{-1}(t)\dot u(t)$, and no additional time dependence.

Now if the equilibrium-contact term in the second line is neglected, Eq.~\eqref{eq:fevo_app} has the same qualitative fixed-point structure as the shear-stress attractor in Bjorken flow. At early times, the drive-dominated part has fixed points at
$f_\Pi=\pm\sqrt{\chi/\E}$; for the monotonic drive in Eq.~\eqref{eq.formofexp}, one of these is attractive and the other repulsive. At late times, the relaxation term leaves a single attractive fixed point at $f_\Pi=0$. The early fixed point is therefore controlled by the scattering-length drive, while the late fixed point is controlled by equilibration.

This idealized analysis identifies the drive-induced attractor mechanism in the near-unitary Fermi gas. The realistic system differs in two ways that are essential for the observable signal: $\chi/\E$ evolves because both $\chi$ and $\E$ depend on the trajectory, and the equilibrium contact term need not be negligible. These effects do not remove the early-time memory-loss mechanism. Instead, they prevent it from appearing as a single universal curve in $f_\Pi(t)$.

\begin{figure}
    \centering
    \includegraphics[width=\linewidth]{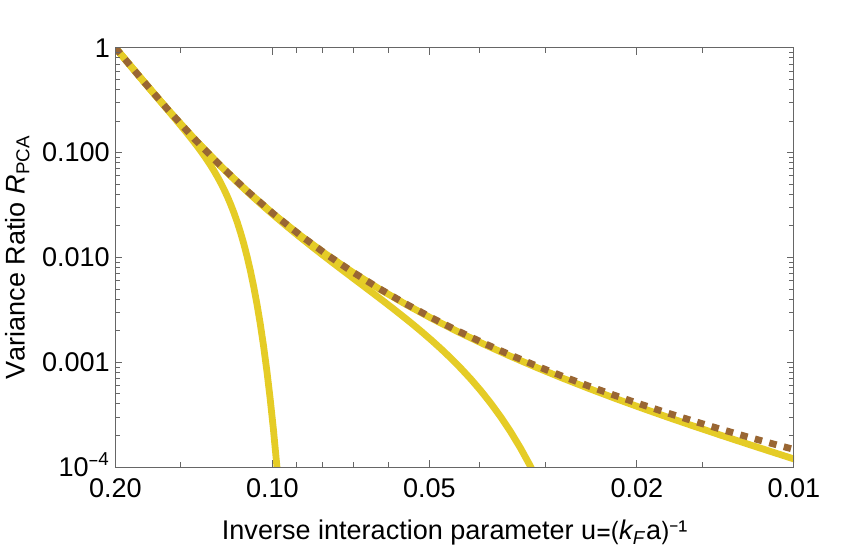}
    \caption{Evolution of the dimensional reduction measure $R_{\rm PCA}$ in terms of the dimensionless scattering length $u$. Shown are the same cases as in Fig.~\ref{fig:pcr}. During stage E/D, the curves agree with the evolution in the absence of transient decay (brown, dashed), and thus also agree with each other. When transient decay sets in, $R_{\rm PCA}$ starts to decay quicker and deviate from the common trend. Faster drives at higher $\alpha$ mean that the system is able to reach lower values of $u$ before this happens.}
    \label{fig:u-independence}
\end{figure}

\section{Drive independence of stage E/D}

After a variable change to $u=(k_Fa)^{-1}$ as the dynamical variable, where the correspondence to time $t$ is given by the chosen drive $a^{-1}(t)$, the form of the evolution equations \eqref{eq:ufevo} and \eqref{eq:uEevo} suggests that the dynamics in stage E/D is independent of the details of the drive and solely depends on the interval swept in $u$. This is verified by agreement with the evolution without transient decay in this stage E/D (see Fig.~\ref{fig:pcr}). The End Matter identifies this as a possible way of experimentally identifying the drive-induced part of dimensional reduction. Fig.~\ref{fig:u-independence} demonstrates more directly that the statement indeed is true for the evolution in our model system.

\section{Equation of state}\label{app:eos}

In the literature, thermal properties and transport coefficients of cold gases are known in terms of their temperature, while time evolution equations consider energy density as the fundamental quantity. This necessitates the establishment of an equation of state to relate the two. For $\theta\gtrsim0.2$, the temperature dependence of the chemical potential $\mu$ in units of $E_F$ and energy density $\E$ in units of $\frac{3}{5}E_Fn$ behave like those of a free Fermi gas shifted by $\xi_n-1$, where $\xi_n\approx0.45$~\cite{Ku_2012}. Thus, the energy density is given as
\begin{align}
     \E&=\frac{3}{5}E_Fn(\xi_n-1)-\frac{3}{(2\pi)^{3/2}}m^{3/2}T^{5/2}\mathrm{Li}_{5/2}(-z)\\&=\left[\frac{3}{5}(\xi_n-1)-\frac{9\sqrt{\pi}}{8}\theta^{5/2}\mathrm{Li}_{5/2}(-z)\right]E_Fn\;,\label{eq:Etz}
\end{align}
which in addition to temperature $\theta$ also depends on the fugacity $z=\exp(\mu_{\rm free}/T)$ in terms of the chemical potential of the free Fermi gas. In order to find the fugacity, we compare the number density of the free Fermi gas
\begin{align}
    n&=-\frac{1}{\sqrt{2}}\left(\frac{mT}{\pi}\right)^{3/2}\mathrm{Li}_{3/2}(-z)
\end{align}

to the same quantity in terms of the Fermi momentum $n=k_F^3/(3\pi^2)$ to relate $z$ to the temperature $\theta$. 
\begin{align}
    \theta^{-3/2}&=-\frac{3\sqrt{\pi}}{4}\mathrm{Li}_{3/2}(-z)\label{eq:tz}
\end{align}
This now fully establishes the relation between energy density $\E$ and temperature $\theta$. However, an analytical expression cannot be obtained, since the polylogarithms $\mathrm{Li}_s$ appearing in~\eqref{eq:Etz} and~\eqref{eq:tz} can not be inverted analytically. Instead, we employ numerical root finding algorithms for a set of temperatures and set up an interpolation routine for $\theta(\E)$.

\section{Temperature dependence of transport coefficients \label{app:transport}}

Close to unitarity, $|k_Fa|\gg 1$, analytical results for the equilibrium contact density $C_{\rm eq}(\theta)$% [cf. Eq.~\eqref{eq:Ceq}]
, the sum rule for the bulk viscosity $\chi(a,\theta)$ and the bulk viscous relaxation time $\tau_\Pi(\theta)$ are known only in the high temperature limit, $\theta\gg1$. In~\cite{Enss:2019ydh}, $\chi$ is given in terms of the thermal wavelength $\lambda_T=\sqrt{2\pi/(mT)}$ and a scattering parameter $v=\lambda_T/(a\sqrt{2\pi})$. In the limit $v\to0$, this evaluates to
\begin{align}
    \chi_{\theta\gg1}(a,\theta)=\frac{2\sqrt{2}}{9}z^2T\lambda_T^{-3}v^2=\frac{16\theta^{-3/2}}{27\sqrt{2\pi}}(k_Fa)^{-2}E_Fn\;.
\end{align}
The same paper also gives a result for $\zeta$ that is of a very similar form, leading to $\tau_\Pi=\zeta\chi^{-1}\propto\theta^{-1}$.

However, in the regime $\theta\gtrsim0.2$ that we are interested in, these results can be quite inaccurate. We instead rely on numerical results for these quantities that were obtained in the Luttinger-Ward model~\cite{Enss:2010qh,Enss:2019ydh}. For $C_{\rm eq}(\theta)$ and $\chi(a,\theta)$, we used interpolations of the numerical datapoints. However, in the case of $\tau_\Pi(\theta)$, our interpolations were quite oscillatory, which was reflected also in simulation results. So instead, we used a fit to the data given by
\begin{align}
    \theta T_F\pi\tau_\Pi(\theta)\approx\mathrm{max}(0.02\theta^{-2.4}+1.8,-2\theta^{-0.4}+3.23)\;.\label{eq:taupi}
\end{align}
For completeness, we also state fits to the data of $C_{\rm eq}(\theta)$ and $\chi(a,\theta)$.
\begin{align}
   \frac{(k_Fa)^2}{ E_Fn}\chi(a,\theta)&\approx\left[100\theta+\left( \frac{(k_Fa)^2}{E_Fn}\chi_{\theta\gg1}(a,\theta)\right)^{-2}\right]^{-1/2}\\
   \frac{C_{\rm{eq}}(\theta)}{4\pi m}\frac{k_F}{E_Fn}&\approx\left[0.42^{-2}+\left(\frac{C_{{\rm eq},\theta\gg 1}(\theta)}{4\pi m}\frac{k_F}{E_Fn}\right)^{-2}\right]^{-1/2}
\end{align}

\end{document}